%% file: main.tex
\documentclass{article}

\input{packages}
\input{macros}
\usepackage{fullpage}
\usepackage[numbers, sort, comma, square]{natbib}
\usepackage{svg}
\usepackage{enumitem}
\usepackage{ifthen}
\usepackage{authblk}
\usepackage{changepage}

\begin{document}

\title{Mayfly: Private Aggregate Insights from Ephemeral Streams of On-Device User Data}


\author[1]{\hspace{-3mm}Christopher Bian}
\author[2]{Albert Cheu}
\author[2]{Stanislav Chiknavaryan}
\author[2]{Zoe Gong}
\author[2]{Marco Gruteser\footnote{Authors are listed in alphabetical order. Correspondence to Marco Gruteser (\texttt{gruteser@google.com}).}}
\author[1]{Oliver Guinan}
\author[1]{Yannis Guzman}
\author[2]{Peter Kairouz}
\author[2]{Artem Lagzdin}
\author[2]{Ryan McKenna}
\author[2]{Grace Ni}
\author[1]{Edo Roth}
\author[2]{Maya Spivak}
\author[2]{Timon Van Overveldt}
\author[2]{Ren Yi}

\affil[1]{\textit{Google}}
\affil[2]{\textit{Google Research}}


\maketitle
\renewcommand{\thefootnote}{\fnsymbol{footnote}}
\makeatletter
\def\@fnsymbol#1{\ensuremath{\ifcase#1\or \dagger\or \ddagger\or
   \mathsection\or \mathparagraph\or \|\or **\or \dagger\dagger
   \or \ddagger\ddagger \else\@ctrerr\fi}}
\makeatother

\begin{abstract}
\input{sections/abstract}

\end{abstract}

\section{Introduction}
\input{sections/introduction}\label{sec:intro}

\section{Background \& Problem Setting}

\input{sections/data_setup}
\label{sec:eie_problem}

\paragraph{\textbf{Actors \& Threat Model}}
\input{sections/threat_model}




\section{Mayfly's Design Overview}\label{sec:design_overview}
\input{sections/design_overview}

\subsection{Streaming DP and Completeness}\label{streaming_dp}
\input{sections/streaming_dp.tex}

\subsection{Programmable Queries}
\input{sections/programmable_queries}

\subsection{Hierarchical Aggregation}
\input{sections/agg_service_design}\label{sec:hierarchical}

\subsection{Aggregation Service}\label{sec:agg_service_impl}
\input{sections/agg_service_impl}

\section{Anonymization Mechanism Design} \label{sec:mechanism_design}
\input{sections/anon_design}

\input{sections/dp_queries}

\section{EIE Case Study Evaluation}\label{sec:eval}
\input{sections/eval_intro}
\subsection{Device Reach and Accuracy}\label{scale_1}
\input{sections/scale_1} 
\subsection{Empirical Evaluation of Mayfly with DP}
\input{sections/dp_eval}\label{sec:dp_eval}

\section{Related Work}
\input{sections/related_work}
\label{sec:relatedwork}

\section{Discussion and Lessons Learned}
\input{sections/discussion}\label{sec:discussion}

\section{Conclusion \& Future Work}
\input{sections/conclusions}

\section{Acknowledgments}
\input{sections/ack}



\bibliographystyle{IEEEtran}

\bibliography{ref}

\appendix
\section{DP Privacy Guarantee}\label{sec:dp_spec}
\input{sections/dp_spec.tex}





\end{document}

%% file: packages.tex
\usepackage{amsmath,amssymb,pifont}
\usepackage{multicol}
\usepackage{amstext}
\usepackage{amsthm}
\usepackage{multirow}
\usepackage{booktabs}
\usepackage[skip=0pt]{subcaption}
\usepackage{times}
\usepackage{lipsum}
\usepackage[shortlabels]{enumitem}
\usepackage{cancel}
\usepackage{wrapfig}
\usepackage{array}
\usepackage{siunitx}
\usepackage{csvsimple}
\usepackage[multidot]{grffile}
\usepackage{bbm}
\usepackage{dblfloatfix}
\usepackage{hyperref}
\usepackage{makecell}
\usepackage{bbm, dsfont}
\usepackage{mathtools}
\usepackage[dvipsnames]{xcolor}
\usepackage{comment}
\usepackage[capitalise,noabbrev]{cleveref}

\usepackage{geometry}
\usepackage{mathpazo}
\usepackage{enumitem}

\usepackage[multiple]{footmisc}
\usepackage{mathrsfs}
\usepackage{todonotes}
\usepackage{tikz}
\usepackage{xurl}
\usepackage{hyperref}
\usepackage{cleveref}
\usepackage{algpseudocode,algorithm,algorithmicx}
\usepackage{xfrac}

\usepackage{graphicx} 
\usepackage{color}

\usepackage{array}
\usepackage{amssymb}
\usepackage{amsmath}
\usepackage{xspace}
\usepackage{fancyhdr}
\usepackage{comment}
\usepackage{nicefrac}
\usepackage{authblk} 
\usepackage{pdfpages}


%% file: macros.tex
\definecolor{orange}{RGB}{227,116,0}
\definecolor{darkgreen}{RGB}{10,116,20}
\newcommand{\gruteser}[1]{[{\color{purple}{\textbf{Marco:} \emph{#1}}}]}

\newcommand{\ryan}[1]{[{\color{orange}{\textbf{Ryan:} \emph{#1}}}]} 
\newcommand{\cheu}[1]{[{\color{cyan}{\textbf{Albert:} \emph{#1}}}]}

\newcommand{\pap}{Mayfly\xspace}

\newcommand{\anon}{0}

\newcommand{\anonymized}[2]{
  \ifthenelse{\anon=1}{#1}{#2}
}

\newtheorem{lem}{Lemma}[section]

\newtheorem{example}[lem]{Example}

\newcolumntype{P}[1]{>{\centering\arraybackslash}p{#1}}
\newcommand{\region}{\texttt{region}\xspace}
\newcommand{\activity}{\texttt{activity type}\xspace}
\newcommand{\direction}{\texttt{trip direction}\xspace}
\newcommand{\distance}{\texttt{distance traveled}\xspace}
\newcommand{\duration}{\texttt{duration}\xspace}
\newcommand{\numtrips}{\texttt{number of trips}\xspace}
\newcommand{\metric}{\texttt{metric}\xspace}

\algtext*{EndFor}

%% file: sections/abstract.tex
This paper introduces Mayfly, a federated analytics approach enabling aggregate queries over ephemeral on-device data streams without central persistence of sensitive user data. Mayfly minimizes data via on-device windowing and contribution bounding through SQL-programmability, anonymizes user data via streaming differential privacy (DP), and mandates immediate in-memory cross-device aggregation on the server---ensuring only privatized aggregates are revealed to data analysts. Deployed for a sustainability use case estimating transportation carbon emissions from private location data, Mayfly computed over 4 million statistics across more than 500 million devices with a per-device, per-week DP $\varepsilon = 2$ while meeting strict data utility requirements. To achieve this, we designed a new DP mechanism for \texttt{Group-By-Sum} workloads leveraging statistical properties of location data, with potential applicability to other domains.

%% file: sections/introduction.tex
The rapid advancement of on-device artificial intelligence (AI) has unlocked new opportunities for privacy-preserving personalization. On-device models can analyze user activities and their surroundings to understand context, as well as directly support users as personal assistants, offering rich and context-sensitive AI experiences. These applications can be kept at a high quality by way of privacy-preserving analytics. One use-case is aggregating direct user interactions with on-device AI models, such as measuring assistant accuracy, user satisfaction, and behavior trends. Another application is computing statistics over the user reports generated by the on-device models.

In this work, we ground in a concrete production use-case: the Environmental Insights Explorer (EIE) \cite{eiewebsite, eieblog}. EIE provides cities with critical information about transportation related greenhouse gas emissions, derived from aggregated and Google Maps Timeline~\cite{mapstimeline} data from opted-in users, which is generated by on-device AI models. This information enables policymakers to understand traffic patterns, identify emission hotspots, and develop effective strategies for reducing their city's carbon footprint.  However, transportation data requires robust privacy protections to prevent the re-identification of individuals or the inference of sensitive attributes. We thus strive to produce aggregate analytics \emph{without centrally storing any individual data}.

Differential privacy (DP)\cite{DMNS} has emerged as the gold standard for computing privacy-preserving statistics. However, in very high-dimensional use cases (where millions of statistics need to be computed), challenges arise when using off-the shelf algorithms, even those that are state-of-the-art. This is especially true when there is a large variance in the magnitude of user contributions. Unlike efforts like the 2020 census \cite{abowd20222020}, which largely computes counts, we compute sums over metrics like ``total distance,'' which can vary dramatically depending on modes of transportation (consider the difference between a user taking a walk in their local park versus taking a cross-continent flight). Because of this structural variance, and because users can contribute multiple rows of information, it is difficult to bound user contribution and scale the resulting sensitivity (and thus the noise applied), in a way that maintains utility of large magnitude metrics without drowning out those of small-magnitude metrics.

To our knowledge, there are no published techniques we could use for this scenario. We operate in the central model of DP (as opposed to the local model, or LDP), since the amount of noise required in the local models requires roughly a factor of $\sqrt{n}$ times higher noise. In our setting with $O(100M)$ users, the amount of noise we end up with is gigantic, and would affect accuracy by a factor on the order of $10,000$. While the local model benefits from the fact that no trust is required in the central aggregator, our system is compatible with using TEEs to ensure end-to-end verification and mitigate insider threats by way of encrypted memory (see \cite{eichner2024, vanoverveldt2025discovering}). 

A standard solution in the central model would jointly clip a user’s entire contribution and add noise. If we were to calibrate the noise to meet our utility requirements, this would result in a privacy parameter $\epsilon > 16$, which does not provide a strong formal privacy guarantee. This is why we apply local activity-level scaling norms: scale data on user devices before their collection, depending on the activities (modes of transportation) taken. We then utilize a single clipping bound across a user’s entire contribution, computed to take into account their re-scaled contributions. These clipped contributions are aggregated securely under federated analytics to a central server, where we perform server-side rescaling (to readjust the data), noise addition (to provide DP guarantees), and an additional post-processing threshold step to discard over-noisy data. We evaluate our algorithm on a data set with hundreds of millions of users, and find that we can achieve a DP guarantee of $\epsilon = 2$ for a privacy unit of (user, week).

This DP algorithm is one component of the system design of Mayfly\footnote{Mayflies belong to the Order Ephemeroptera, derived from the Greek words "ephemeros" meaning "short-lived" and "pteron" meaning "wing." This name reflects their remarkably brief adult lifespan, which typically lasts from a few hours to a few days. They are also known for their swarming behavior.}, which achieves privacy, utility, and scale through three central pillars: (1) lightweight, programmable on-device data minimization, (2) immediate ephemeral aggregation, and (3) a streaming differential privacy mechanism. \pap allows for specifying the desired results via a SQL interface, restricted to summations across many user devices. Queries containing per-client logic are downloaded to devices, minimizing data by only selecting and summarizing relevant information. This results in more focused updates streamed to servers, with no potential identifiers linking to a specific device or user (aside from the data being aggregated and necessary network identifiers like the device's IP address). Given the time-series nature of on-device user data, the platform supports queries over predefined upcoming time windows or streaming data with tumbling windows. Server-side aggregation logic then operates ephemerally: we perform in-memory aggregations and individual contributions are not accessible afterwards. Finally we apply DP to ensure that released aggregates do not overfit to any single user contribution. Despite stringent privacy constraints, the platform is designed to maintain high fidelity in aggregate results. This is crucial when computing statistics over billions of devices, particularly for statistics derived from diverse user activity.

\paragraph{\textbf{Key Challenges}} Several complexities arise when using differential privacy in a federated, streaming context. 

\begin{itemize}  
    \item \textit{Minimizing Noise-Induced Error:} DP requires noise addition, which can degrade result accuracy. Minimizing the impact of noise added for privacy through careful mechanism design is crucial to maintain data utility for downstream analyses.
    \item \textit{Data Streaming and Completeness:} Streamed data is aggregated over pre-defined time windows (TWs), which are aligned with differential privacy units. This requires devices to maintain records to ensure that each time period is complete before they perform any local contribution bounding or upload data.
    \item \textit{Device Heterogeneity:} With billions of smartphones worldwide, varying device capabilities impact participation in computations. While there may be ways to ensure ephemeral server-state by offloading certain aggregations to devices, this also comes with downsides. The more work required of devices, the more likely it is that less powerful or less well-connected devices are under-represented, potentially biasing results.
\end{itemize}

\paragraph{\textbf{Our Contributions}} In summary, the main contributions of this work are:

\begin{itemize}
    \item A new DP mechanism (\cref{sec:mechanism_design}) for \texttt{Group-By-Sum} workloads built on top of the above system, which is particularly effective for location data.  For a target utility of $3\%$ relative error, our new mechanism improves privacy budget consumption by over $8\times$ as compared to existing baselines (which represents a corresponding exponential decrease in the likelihood of data leakage).
    \item An end-to-end system design (\cref{sec:design_overview}) that supports distributed SQL queries, enforces early data minimization on-device, maintains ephemerality, enabling us to never store individual data on a central server, and can stream outputs respecting a privacy time unit through a latency-tolerant design.
    \item An evaluation (\cref{sec:eval}) and lessons learned (\cref{sec:discussion}) from a large-scale production deployment of the system to perform a DP data release across 500+ million user devices. This includes a design of lightweight workloads, native group summation, iteration on device constraints, and more efficient task creation and assignment to increase our device reach from $49\%$ to $93\%$ from a baseline approach.
\end{itemize}

\paragraph{\textbf{Related Work}}\footnote{Note that this related work is an abridged overview; we have a more complete discussion in ~\cref{sec:relatedwork}.}
There is a great body of previous work related to large-scale private data aggregating systems. Works like \cite{roy2010airavat} and \cite{mcsherry2009privacy} similarly handle SQL queries, but only in a server-side setting, meaning they have access to all data in central storage, which we aim to avoid. Large-scale systems like \cite{corrigangibbs2017}, \cite{roth2019honeycrisp}, and \cite{froelicher2017unlynx} have a different set of actors, relying on either a 2-server model or a committee of devices/servers to distribute trust. These are not suited for our use case, where we are not able to rely on an external party to assist in processing sensitive transportation data.

There are other works that rely on cryptographic primitives (see \cite{mansouri2023sok} for a survey) or TEEs (\cite{wang2022panda}) to achieve their privacy goals, resulting in additional client overhead, which are challenging for our low-resource devices to support. Other generic approaches to answer large-scale aggregation queries, such as OLAP~\cite{chaudhuri1997overview}, do not address privacy, and could allow service providers to view partial slices or aggregates of user data.

Therefore, we believe this work is a unique contribution that shows how to minimize client-side data and avoid storing individual data centrally, while making computation feasible over hundreds of millions of devices of varying resources, and supporting strong central DP in a single-server setting.

%% file: sections/data_setup.tex
\paragraph{\textbf{Data}}

We consider a federated data model where each user device produces a collection of records, which in general can be arbitrary but for sake of concreteness will typically be tabular in nature and contain a mixture of categorical attributes and numerical attributes.

 
\begin{example}[\anonymized{Transportation}{Timeline} Data]
\label{ex:timeline_data}
Each user record has three categorical attributes: \region (e.g., New York City, Tashkent, Santiago), \activity (e.g, cycling, driving, subway), and \direction (within region, outbound, and inbound).  There are $\approx 50,000$ possible regions, 9 activity types, and 3 trip directions.  Each record also has two numerical attributes: \distance (measured in kilometers) and \duration of the trip (measured in seconds).  Each record is timestamped on-device, and there are on the order of hundreds of millions of users contributing billions of trips each week.
\end{example}

\paragraph{\textbf{Privacy Goals}} Privacy is inherently multifaceted, and there are several complementary privacy-preserving principles we aim to incorporate into the design of \pap.

\begin{enumerate}
     \item \emph{Data Minimization}: The server should process only the minimum amount of information required to answer the query. As one result of this goal, we require that user contributions should be appropriately bounded \emph{before} aggregation on the server.
    \item \emph{Cross-Device Aggregation}: An analyst querying the system is limited only to a specific class of Group-By-Aggregates. These Queries must enforce aggregation---the analyst should only have access to  highly aggregated data across many user devices.
    \item \emph{Ephemerality (Data Retention)}: The system should retain data for the minimum amount of time necessary at all levels---both on device and at the server. When possible, it should prefer processing in-memory over persisting data. Data should be automatically deleted from devices after specified time windows.
    \item \emph{Non-targetability}: The system should not allow targeted queries aimed at obtaining information from a particular device of interest or queries aimed at identifying a device that with particular data of interest. Its API should be limited to allow only a specific class of less sensitive aggregate queries.
    \item \emph{Differential Privacy}: Artifacts derived from on-device user data are only released from the system if they satisfy a suitable device-level differential privacy (DP) guarantee \cite{DMNS}.\footnote{This guarantee is not strictly enforced by \pap, although it is present in the case study we present for \anonymized{Product A.}{EIE.}}  
\end{enumerate}

The final goal of differential privacy requires a more formal treatment. That is, we must specify the unit of information that should be protected via an appropriate adjacency relation between datasets.  In our setting, a user generally contributes data over time so it is meaningful to define adjacency in the following way: for any time window $w$ that defines a complete dataset of device records, $D_w$ and $D'_w$ are adjacent when $D'_w$ can be formed by adding or removing one devices's records to $D_w$. Thus, the \emph{privacy unit} is (device, window) addition.

Now that we have defined the privacy unit, we define the DP objective. For any pair of adjacent datasets $D_w,D'_w$ and any given event $E$, an $\epsilon$-DP mechanism $\mathcal{M}$ guarantees the odds of the event changes by at most an $\exp(\epsilon)$ factor, that is:
\begin{equation}
\mathrm{Pr}[\mathcal{M}(D_w) \in E] \leq \exp(\epsilon)\cdot \mathrm{Pr}[\mathcal{M}(D'_w) \in E] \label{eq:dp}
\end{equation}

Differential privacy is the gold standard for private data analysis, as it enables learning about the population as a whole (through the mechanism $\mathcal{M}$), with a strong and quantifiable impact on the privacy of the individuals in the population (via \cref{eq:dp}).

Simultaneously satisfying all of these privacy goals while adhering to practical device constraints is a challenging problem; we detail how we overcome these challenges in \cref{streaming_dp} and \cref{latency_tolerant}.

%% file: sections/threat_model.tex
Our system involves four groups of actors, summarized in \cref{fig:dp_agg}.
A \emph{data analysis} team issues a query to our system on behalf of a \emph{data consumer}.
Client \emph{devices} send data to servers (which have \emph{server administrators}). These servers produce private outputs for the data analysts, who pass it along to the consumer.

Each actor could be adversarial; we overview their assumed capabilities now, beginning with client devices. Devices naturally come in contact with raw personal data, so a compromised device could clearly violate privacy. Mitigations for this threat, such as physical security and software updates, are out of scope for this work.

We move on to the server administrators. Because standard precautionary measures are in place, such as tight access control policies and dual control\footnote{Dual control, also known as two-person control, refers to the policy where a proposed change can only be made after approval by someone other than the proposer \cite{anderson2020security, nistTPC}.}---an adversarial server administrator is effectively rendered passive, in that they can monitor but not modify computations.\footnote{Trusted execution environments could strengthen the confidentiality guarantee, but we leave this for future work.} We also assume that they cannot observe ephemeral server-side state, like the contents of memory or partially aggregated results. Hence it suffices to limit persisted server-side state to aggregates.

The data consumer's point of contact with the system is via the analysis team, so we focus on threats by the latter. For example, an adversarial analyst could attempt to issue queries that target the past locations of individuals. This includes adversaries who are curious about specific individuals and would like to learn the locations that they visited, but also adversaries who are curious about a specific location and would like to learn the identities of users that visited this location in a certain time interval. We defend against this threat both through our privacy goal of \textit{non-targetability}, as well as by answering queries subject to differential privacy; refer to the privacy goals above.

%% file: sections/design_overview.tex

\pap's architecture includes a client interface that allows data analysts to make streams of data available to private aggregations, a SQL-based server interface for data engineers to express privacy-preserving queries over data streams, hosted services for distributing queries to devices and aggregating device results, and anonymization pipelines. To enable use in a variety of personal AI contexts, \pap was designed to remain compatible with the federated learning infrastructure foundation~\cite{bonawitz2019}, thus enabling both learning and analytics insights tasks over the same data. In this paper, we focus solely on the analytics design. 

To support DP in a streaming setting, with new data points constantly created, we configure our computations around the notion of a \textit{time window (TW)}, which represents the window of time over which the DP guarantee applies, such that the privacy unit is (device, TW). \pap queries are continuous queries that are conceptually executed over each TW to produce one DP aggregate for each time window. To achieve this, the privacy time unit is reflected in multiple aspects of our system design, including (1) the client windowing configuration, (2) the query implementation, and (3) the server aggregation configuration.

\begin{figure*}[t]
  \centering
  \includegraphics[width=\textwidth]{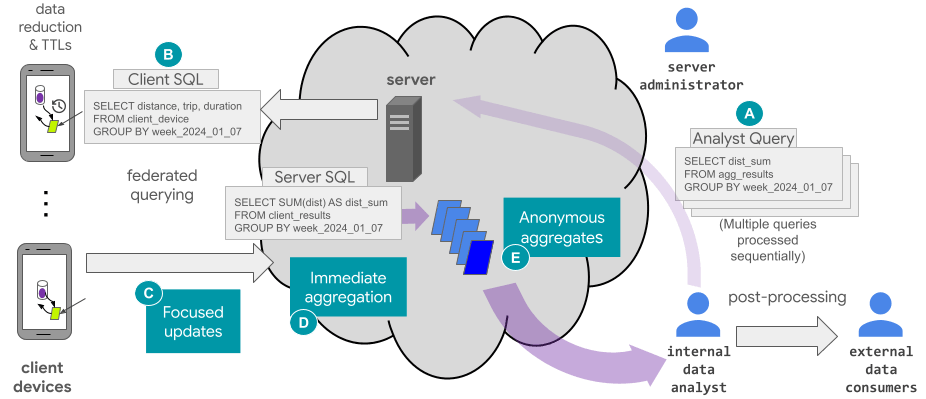}

  \caption{Overview of all system components and actors. Analyst queries (A) are converted into lightweight SQLite-compatible subqueries for execution on device (B). This includes some DP parameters, and allows devices to locally bound their contributions, and upload focused updates (C). The federated server performs an immediate aggregation (D), resulting in server-side aggregates that have a strict time-to-live. Anonymization (E) adds noise and prepares the data for internal and external release, with some optional post-processing to further minimize the data and give focused results for external consumers.} \label{fig:dp_agg}
\end{figure*}

We begin with an overview by describing the lifecycle of a query as depicted in \cref{fig:dp_agg} and how this design addresses the privacy design goals.

\begin{enumerate}[label=\Alph*.]
    \item {\bf Analyst Query.} Analysts submit a task specification, which includes a continuous \pap query, to the system with two-person control (code review and approval by a second engineer). The specification is then converted into (i) a client task  with a SQLite subquery and (ii) a configuration for the server to perform cross-client aggregation and anonymization for this query. The server makes the task available to clients that request work. This task/query must aggregate data over a series of windows in time occurring after the submission of the query to the system; that is, retrospective queries (aggregating the last weeks data) are not supported.
    \item {\bf Task assignment.} Clients check into the server at their convenience to request client tasks to execute. The timing of check-ins can be adjusted based on data availability and device resource consumption tradeoffs. For example, clients may be configured to contact the server at most once per day, when the device has remained idle for an extended period and network connectivity is available. If tasks are available, the server assigns one or multiple tasks to clients for download and execution. To save resources, each task can optionally include a lightweight eligibility pre-computation that clients will execute to determine which of the tasks are worth executing (to determine if new, relevant data for the task is available, for example).
    \item {\bf Focused updates.} Clients execute eligible tasks over a window of event records. This performs {\em data minimization} on-device before sending, and transforms data into a format compatible with cross-device aggregation and differential privacy. Clients then send the minimized and transformed data to the server.
    \item {\bf Immediate Aggregation.} The server, which has been configured to execute the mandatory {\em cross-device aggregation} part of the query, executes a hierarchical aggregation pipeline that progressively combines device results into highly aggregated results as they arrive. Its first stage, which we also refer to as aggregation service, performs immediate in-memory aggregation over batches of device results, before caching any data on disk. This ensures {\em ephemerality} of individual records. We further minimize retention of and restrict access to these partial aggregates, letting cache retention increase commensurate with the level of aggregation.
    \item {\bf Anonymous Aggregates.} The final stage of the aggregation pipeline is a latency-tolerant anonymous release. After allowing straggler clients sufficient time to contribute, the release mechanism combines all partial aggregates for a TW, applies the noising step of a {\em differential privacy} mechanism, and makes the result available to the analyst.
\end{enumerate}

{\em Non-targetability.}\label{sec:guarding} 
To guard against invasive queries, the design provides no ability to target a query at specific users. Queries are executed on batches that are constructed based on device connectivity and usage constraints. Assuming that satisfying these constraints is a random event hidden from an adversary, this means there is no guarantee that an individual device will be included in the query. The system also offers no access to user or device identifiers. It further structurally requires that queries include a cross-device aggregation clause, so that only cross-user aggregates can be produced. This restriction is also enforced by having a purpose-designed native implementation of grouped summation that translates the original query and performs the cross-user aggregation step (described in more detail in \cref{sec:agg_service_impl}). Finally, devices provide access only to the current stream of evolving on-device data---this implies that it is impossible to access past data or to reactively query the stream at any point in time based on the results of a prior query.

{\em Device Heterogeneity and data utility}. One important goal of the platform is to gather accurate statistics over a large fleet of heterogeneous devices that have very different connectivity patterns, data distributions and so on. Reusing the federated learning infrastructure described in ~\cite{bonawitz2019} would mean the system might be subject to bias due to powerful phones receiving preferable selection for each round. This analytics focused design addresses this by using computationally lighter-weight SQL tasks, optimized task selection and deployment (\cref{sec:taskselection}), and the latency-tolerant release design (\cref{latency_tolerant}).

%% file: sections/streaming_dp.tex
While Section \ref{sec:mechanism_design} explains the DP mechanism in full, our DP implementation boils down to two components, applied to each privacy time unit: (1) bounding the amount that each client can contribute to any released statistic, and (2) adding carefully calibrated noise to the aggregates.

\begin{figure}
  \includegraphics[width=\columnwidth]{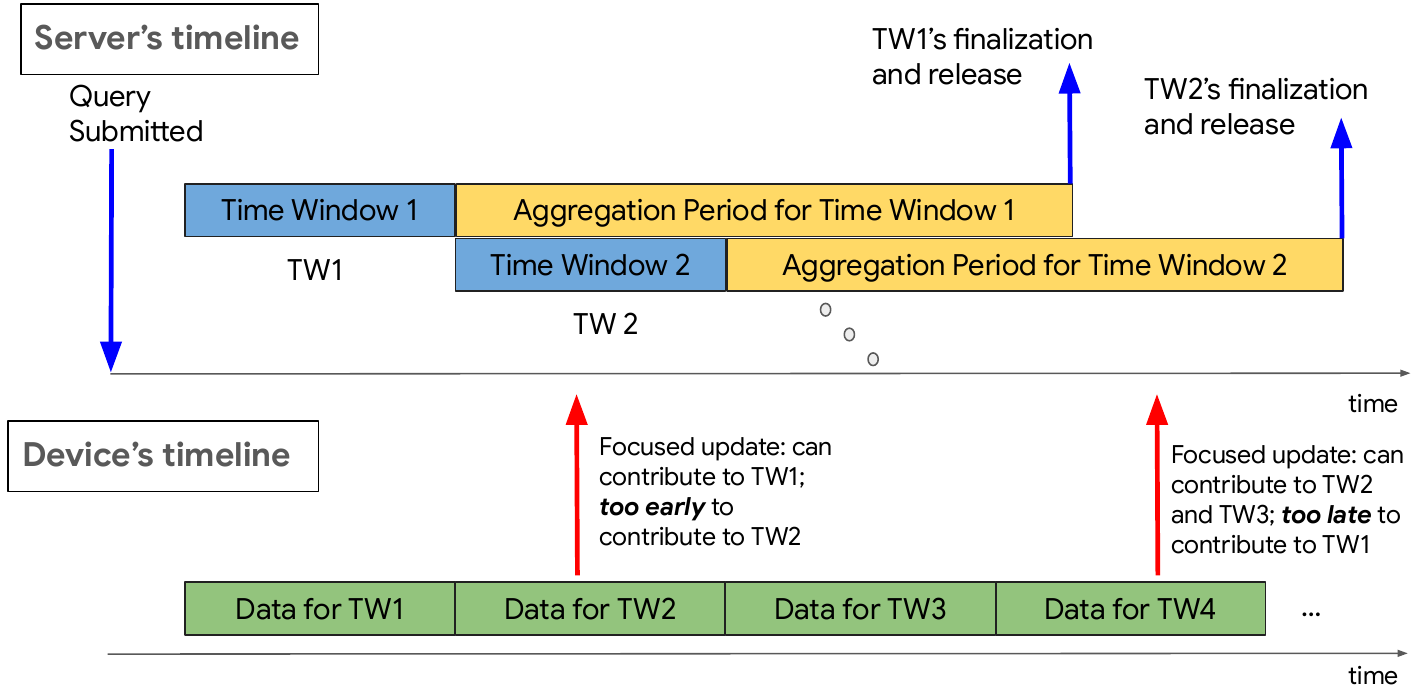}
  \caption{Server and device timelines for multiple time windows.} \label{fig:tw}
\end{figure}

We focus here on step (1), which we refer to as \textit{contribution bounding}. To correctly achieve DP properties, we need to guarantee that each device should only contribute to each privacy time unit at most once, and that the contribution bounds for each device apply over the full time window. There are thus two reasons for windowing and contribution bounding to happen \textit{on device}, as opposed to server-side. First, this enables us to uphold our principle of early data minimization, avoiding uploading any data that is not strictly necessary. Second, this creates less contribution state that needs to be tracked server-side. \cref{fig:tw} demonstrates how we handle streaming contributions to multiple time windows.

 \paragraph{\textbf{Server perspective}} The analyst submits a query before the start of the first time window (TW1). After the window ends, there is an aggregation period for which the server continues to collect and aggregate contributions for this TW1. At the end of this period, all data aggregated over TW1 will be noised and released. Simultaneously, data for TW2 is still being aggregated and prepared for release.
 
  \paragraph{\textbf{Device perspective}} Once the device receives a query, it can start contributing data to appropriate time windows. The red arrows demonstrate the time when a device sends a focused update---note that before TW2 ends, the device cannot contribute their data to this window, but they can still contribute data from TW1, since this falls in TW1's aggregation period. After TW1's aggregation period ends, the device can no longer contribute to this window, but it can contribute to TW2 and TW3 in one focused update. These contributions will appear in aggregate releases in the future.
  
  To implement this behavior on-device, each device maintains an event cache and two watermarks per active query: a high watermark for all data that has already been processed and contributed, as well as a low watermark that advances when system time reaches a new increment of the TW. Only the cached records between the two watermarks will be available to the query upon next execution. Once the query results are sent for aggregation, the high watermark advances to the low watermark (its value at query processing time).

The high watermark ensures that the same event data cannot be contributed more than once to an aggregate defined by a query. The low watermark effectively delays contribution until one or more full TWs of data have accumulated and ensures that the query logic (and contribution bounding) operates over complete TWs.  In practice, it is often convenient to align TWs with calendar days, weeks, or months. We therefore implemented the mechanism by rounding the current time down to such a time boundary. For example, the duration and alignment can be expressed as follows.
\begin{verbatim}
end_time_boundary: {
  time_adjustment:
    ROUND_DOWN_TO_CIVIL_WEEK
}    
\end{verbatim}

We limit retention of data in the on-device storage through an explicit time-to-live (TTL) mechanism that purges a datum from the storage when the difference between the system time and its event time exceeds the TTL setting (e.g., a few weeks). This ensures that data is not retained for extended periods of time, though it can incur trade-offs in query accuracy, since some data may be lost).

Once collected and aggregated across devices, the server releases data at the granularity of a TW, covering aggregations across each one separately.


%% file: sections/programmable_queries.tex
The \pap task specification that analysts submit includes the following components: a \pap SQL query, a data stream selection, a privacy time unit setting, an anonymization configuration, an optional eligibility computation. It also includes additional task metadata prescribing the device (sub)population that the task can execute on, and some limits on how frequently the task should execute on device. 
Here we focus on the SQL query, which contains two subqueries: a client query that is executed in parallel on all devices and a server aggregation query that specifies how the server should aggregate data across devices. 

The client query enables analysts to minimize data collected from devices. Devices execute the query over a table of event records with one or more complete privacy time units as provided by the windowing mechanism (Section~\ref{streaming_dp}). This ephemeral table is created in an
in-memory SQLite database each time a query execution begins. Each event record contains an event time stamp and one or more data fields describing the event. For simplicity, we assume that the windowing mechanism also annotates each event record with a privacy time unit identifier, such as a string representation of rounding date. The client query must produce outputs grouped by the privacy time unit. Devices can also produce multiple event streams and the streams available to the query are defined in the data stream configuration.

The server query must specify an aggregation across devices. Conceptually, the server query is executed over a virtual table that represents a union of all client results. Note though that the system never materializes this table. In addition, the SQL specification is only for analyst convenience since it is never executed on a SQL engine---instead the server query is translated into a configuration for a cross-device aggregator (\cref{sec:hierarchical}).


\begin{example}
Consider the example queries in ~\cref{alg:splitwindowedsqlquery}, which operate over the trip event records from ~\cref{ex:timeline_data}. The client query first sums the travel distances across all individual trips taken for each region and privacy time unit (on each device) to minimize the data sent to the server. The cross-client aggregation query then computes an aggregate travel distance across device for each region and TW.
\end{example}

\begin{algorithm}[t]
\caption{Distributed SQL query example}
\label{alg:splitwindowedsqlquery}
\begin{algorithmic}
    \State \textcolor{blue}{\textbf{SELECT}}  \Comment{Client Query}
    \State \indent \textcolor{orange}{region},
    \State \indent \textcolor{orange}{privacy\_time\_unit},
    \State \indent\textcolor{blue}{\textbf{SUM}}\textcolor{blue}{(}\textcolor{orange}{trip\_distance}\textcolor{blue}{)}\textcolor{blue}{\textbf{ AS}} \textcolor{orange}{user\_trip\_distance}
  \State \textcolor{blue}{\textbf{FROM}} \textcolor{teal}{DeviceDataStream}
  \State \textcolor{blue}{\textbf{GROUP BY}}
  \State \indent \textcolor{orange}{region},
  \State \indent \textcolor{orange}{privacy\_time\_unit}
\State \textcolor{blue}{)}
\State
\State \textcolor{blue}{\textbf{SELECT}} \Comment{Cross-Client  Aggregation on Server}
  \State \indent \textcolor{orange}{region},
  \State \indent \textcolor{orange}{privacy\_time\_unit},
  \State \indent \textcolor{blue}{\textbf{SUM}} \textcolor{blue}{(} \textcolor{orange}{user\_trip\_distance} \textcolor{blue}{)} 
\State \textcolor{blue}{\textbf{FROM}} \textcolor{teal}{UserResults}
\State \textcolor{blue}{\textbf{GROUP BY}} \textcolor{orange}{region}, \textcolor{orange}{privacy\_time\_unit}\textcolor{blue}{;}
\end{algorithmic}\end{algorithm}

Analysts control what on-device data is supplied to the SQLite database, as well as a time range for input data to the per-user query. However, only data generated after the \pap task is launched is available. Within minutes of a \pap task being registered, the client query is received by the first user devices, and results of that query are returned to the server within hours of task registration.



%% file: sections/agg_service_design.tex

On the server, a hierarchical aggregation pipeline processes device contributions to ephemerally generate cross-device aggregates. That is, it immediately sums results across devices in memory, in contrast to conventional systems which first log the received device contributions to disk before invoking analysis pipelines. Note that at the scale of hundreds of millions of device results per query, there exists a tradeoff between the robustness to data loss and the degree of ephemerality achieved. A failure in an aggregation host could lead to a loss of a large portion of the data, with little ability to recover since the original data may no longer be available (in conventional systems, an analysis pipeline can simply be re-run over the data logs). Additionally, the processing pipeline has to accommodate both bursts in arriving device contributions and straggler devices who may contribute days later. 

We address this through an aggregation hierarchy where retention durations increase as data is progressively more aggregated: at the lowest level thousands of individual records are immediately processed into partial aggregates fully in-memory via an aggregation service. These less sensitive partial aggregates are then cached in temporary protected storage (encryption-at-rest, automated deletion based on time-to-live, no human users with access permission) as checkpoints, say for a few days. At the next level, the partial aggregates are then combined through a periodic batch pipeline into larger aggregates over millions of devices, which are cached in temporary protected storage for, say, a few weeks. At the highest level, the final aggregate is computed from these larger aggregates, and differential privacy noise can be applied.

\subsection{Task Selection and Deployment}
\label{sec:taskselection}
When a federated task is registered, the federated service coordinates continuous ephemeral aggregation. The service publishes the task configuration to the cloud where it is accessible to clients, and then creates a new aggregation session in the Aggregation Service. The federated service is now ready to assign clients to the task. 

To minimize client resource consumption and user disruptions, the design allows clients to determine when to process queries based on criteria such as device idle status, network connectivity, and battery level. To signal their availability for query processing, clients check in periodically with the federated server based on the aforementioned criteria, timing hints provided by previous check-ins, and the configuration specified for a given query. For example, the server may suggest client check ins at most every 24 hours. When a client first checks in, the federated server provides a list of eligibility criteria for each registered task in the population. The client determines whether it is eligible for each task (e.g. the client has relevant new data to add or the client has not participated in the task for a minimum configured amount of time) and returns this information to the server.

\begin{figure}
  \centering
  \includegraphics[width=\columnwidth]{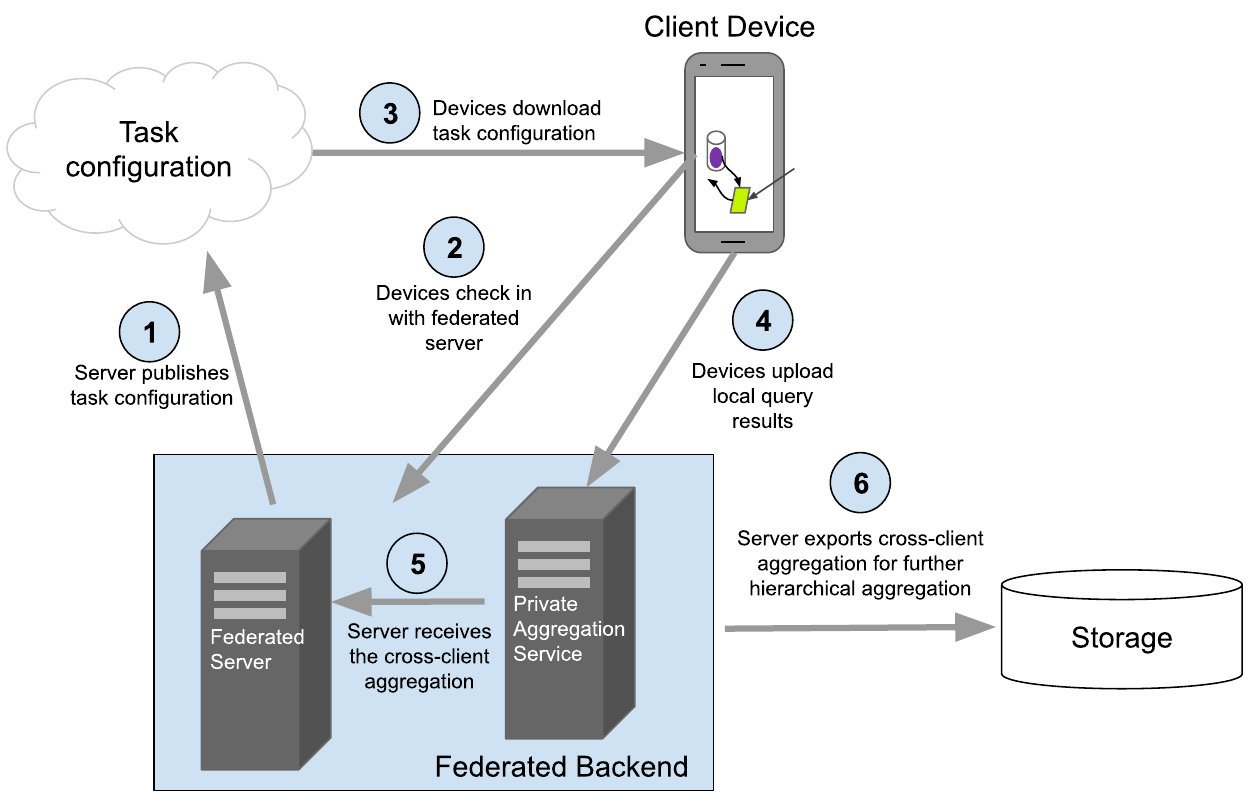}
  \caption{Task assignment architecture and immediate aggregation workflow. \label{fig:agg_flow}}
\end{figure}

The server then assigns the client a list of tasks that it should participate in, based on which tasks have an available active session for which the client is eligible. Each task in the list contains information about where the client can download the SQL query for the task, and which aggregation session the client should upload the results to. The architecture of task assignment and the immediate aggregation workflow is shown in \cref{fig:agg_flow}.

\subsection{Latency-tolerant Anonymous Release}\label{latency_tolerant}

There exists a long tail of heterogeneous devices with limited network connectivity, which cannot execute queries in a timely fashion, especially when the aforementioned device resource consumption optimization are active. This imposes a classic three-pronged tradeoff: choose two among the goals of device resource consumption, latency, and result accuracy. Computing statistics based on a sample of devices that happened to contribute quickly optimizes for latency and device resource consumption but leads to biased results. This is because these devices tend to be more capable and have better network connectivity, which correlates with users from higher socioeconomic status. Programming devices to make more frequent attempts to contribute, in order to achieve unbiased accurate results quickly, would increase device resource consumption. Thus, given that the results for most analytics task such as\anonymized{Product A}{EIE} are not very latency sensitive, we choose instead to optimize for device resource consumption and result accuracy.

This leads us to a latency-tolerant design that allows devices to still contribute data to past TWs if they have been unable to contribute before (due to network connectivity or other constraints). Devices are only limited by the time-to-live setting of their own cache in how far these contributions can extend into the past. The lower and intermediate stages of the aggregation pipeline produce partial aggregates for each TW.

The final stage of the aggregation pipeline is the release stage, which aims to produce a single final release for each TW. The release stage for a TW is triggered when a grace period has passed after the end of the unit---that is, when the system time becomes greater than the sum of the grace period and the TW's ending time. Any data arriving after the release stage has triggered will be ignored and discarded. The grace period configuration therefore defines the latency-tolerance of the pipeline.

The release stage is also responsible for the noising part of the differential privacy mechanism. Specifically, it applies Laplace noise and scaling transformations to each aggregate value in the result. We will detail this further in \cref{sec:mechanism_design}.

%% file: sections/agg_service_impl.tex
The aggregation service implements immediate ephemeral aggregation, the first stage of the hierarchical aggregation process, as a generalized horizontally scalable service. To achieve scale needs, the frontend routes query results from clients to multiple concurrent aggregation sessions on a pool of backend nodes, choosing placement of each session to load balance between the backend nodes. Each aggregation session is tied to a specific query --- when analysts submit a new task, the federated server notifies the aggregation service to create corresponding aggregation sessions.

Individual client inputs are processed and accumulated ephemerally, that is handled only in-memory, until the configured privacy criterium for reporting the results to the next stage of the hierarchical aggregation process is reached. As a privacy criterium, the aggregation service supports a threshold on the minimum number of client contributions, say thousands to tens of thousands, and can be extended with other criteria. In the current system implementation, the federated server controls start and termination of sessions: each session produces status updates that are polled by the federated server, and based on that information the server determines whether to complete a session or forward more clients to it.

Inside the aggregation sessions, the aggregation process is delegated to \textit{Aggregation Cores}. An aggregation core is an abstraction of an aggregation process and its ephemeral state, as shown in \cref{alg:privateagg}.

\begin{algorithm}
\caption{Aggregation core interface}
\label{alg:privateagg}
\begin{algorithmic}
  \State \textbf{AggregationCore $ < $ Params, T, R $ > $}:
  \begin{itemize}
    \item \textbf{Init(p: Params)}. Constructs new core with zero state, p defines privacy threshold.
    \item \textbf{Accumulate(t: T)}. Adds a single contribution to the state.
    \item \textbf{Merge(other: AggregationCore $ < $ Params, T, R $ > $)}. Combine with the state of another core.
    \item \textbf{CanReport() $ \rightarrow $ bool}. Returns true if accumulated value can safely be reported, given initial params.
    \item \textbf{Report() $ \rightarrow $ R}. Returns accumulated value.
  \end{itemize}
\end{algorithmic}
\end{algorithm}

To achieve generality and efficiency, the aggregation core accepts client results in a custom binary format consisting of key-value pairs, with string-typed keys and tensor-typed values which can be summed efficiently. For \pap each output column of a client SQL query is represented by a 1D tensor. However, aggregation cores can also be used to aggregate multi-dimensional values for federated learning use cases.


Client inputs are fully ephemeral throughout the process---once a client input has completed its upload to the Aggregation Service frontend, it is promptly forwarded to the correct backend node that hosts the target aggregation session, where it is immediately aggregated. Thus, the system has fairly low memory requirements and a great memory locality. For any ongoing aggregation session, it needs only to store in-memory a running aggregate of accumulated client inputs, plus a few in-flight client inputs that are awaiting aggregation. The high-performance implementation of aggregation algorithms ensures sufficient throughput to handle incoming client inputs on the fly without queuing. For a typical \pap\anonymized{Product A}{EIE} task, end-to-end aggregation of a single client input takes between $200 \mu s - 400 \mu s$. Considering partial parallel processing on concurrent inputs, a single aggregation session can handle up to $5,000-10,000$ client input aggregations per second.

%% file: sections/anon_design.tex
There are many ways one can incorporate differential privacy into our platform, and the platform itself is general enough to support many different mechanisms.  The right mechanism to use ultimately depends on the workload, or the queries the platform needs to answer with differential privacy.  In this section, we discuss a common scenario where the workload contains a collection of \texttt{Group-By-Sum} queries.  

\begin{example}[\anonymized{Product A}{EIE} Workload] \label{example:workload}
For the\anonymized{Product A}{EIE}application, we would like to release three aggregate metrics: total \numtrips, \distance, and \duration, broken down by \region, \direction, and \activity.   This amounts to releasing three massive histograms (one per metric), each containing $50K \text{ regions} \times 3 \text{ trip directions} \times 9 \text{ activity types}= 1.35$ million partitions.  These statistics are post-processed by applying vehicle fleet distributions and their fuel carbon intensities in order to estimate CO$_2$ equivalent emissions per region, then surfaced to\anonymized{Product A}{EIE} for further use.
\end{example}

One can answer Group-By-Sum queries with differential privacy in many different ways.  One natural baseline is to simply apply the Laplace mechanism \cite{DMNS}: given a vector-valued contribution for each user $\mathbf{v}_1, \dots, \mathbf{v}_n$, and an $L_1$ clipping bound $C$, the Laplace mechanism releases 
$$\mathbf{\tilde{v}} = \sum_{i=1}^n \text{clip}(\mathbf{v}_i, C) + \text{Lap}\Big(0, \frac{C}{\epsilon}\Big)$$

where $\text{clip}(\mathbf{v}_i, C) = \mathbf{v}_i \cdot \text{min}(1, \nicefrac{C}{|| \mathbf{v}_i} ||_1)$.
This mechanism is known to provide $\epsilon$-DP. $C$ is a crucial hyperparameter that must be carefully chosen to balance the bias from clipping and the variance due to Laplace noise.  We know from prior work that choosing $C$ based on an appropriate quantile (e.g., 95\%) of $|| \textbf{v}_i ||_1$ typically provides good utility \cite{liu2023algorithms,andrew2021differentially}.  This quantile can be calculated privately using e.g., the exponential mechanism.  Alternatively, $C$ can be tuned via a grid search using proxy data which will not require DP protections.  There are two natural ways one can use the Laplace mechanism to answer \texttt{Group-By-Sum} workloads.

\begin{figure*}[t]
\includegraphics[width=\textwidth]{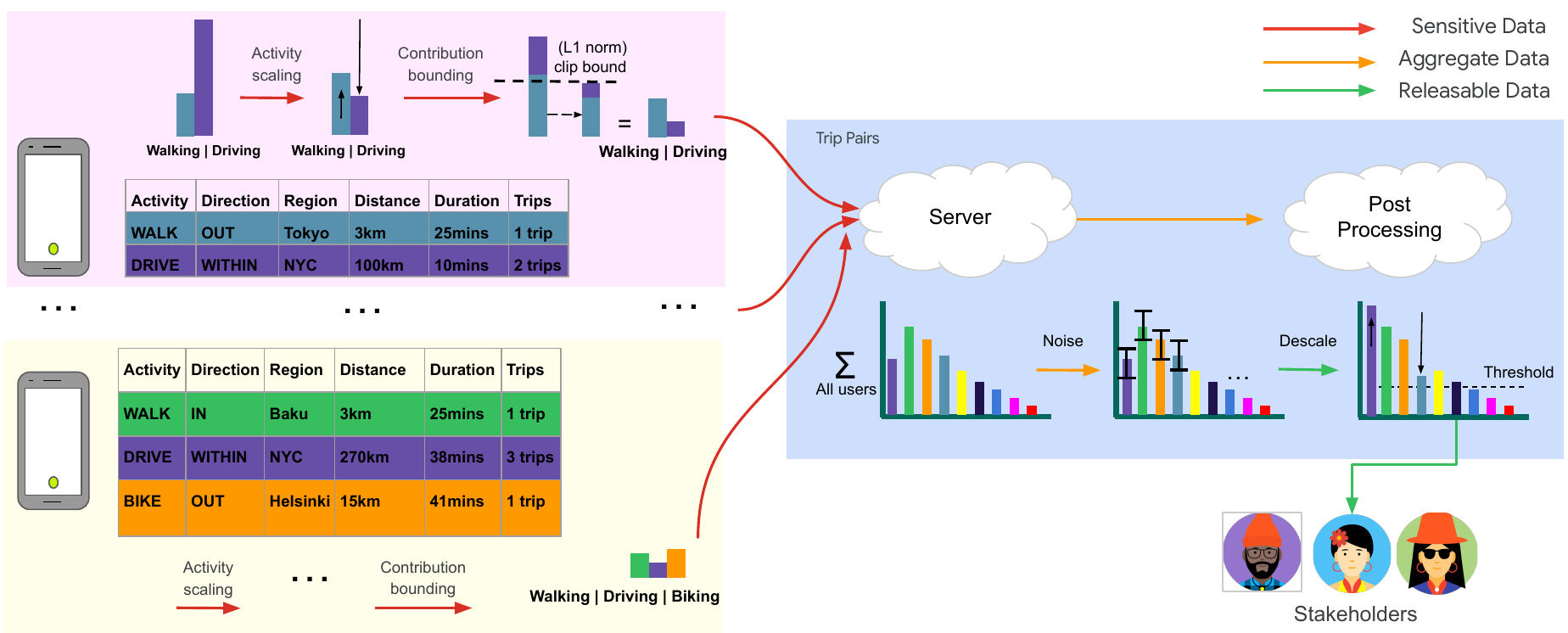}
\caption{A complete overview of the data collection and processing steps, including the ``Activity + Metric Scaling Mechanism'' (with two sample devices shown).  Data starts on device, is scaled and clipped locally, before being aggregated hierarchically on the server. Once aggregated, we add Laplace noise centrally, and perform post-processing steps of descaling and thresholding, before releasing the private data to downstream emissions calculations and then to relevant stakeholders.} \label{fig:mechanism}
\end{figure*}

\begin{enumerate}
\item \texttt{Joint Clipping.} In general, we could take $\mathbf{v}_i$ to be the joint histogram over all groups and metrics.  The main drawback of this approach is that the Laplace mechanism adds i.i.d. noise to each entry of $\mathbf{v}$, and if the entries of $\mathbf{v}$ have vastly different magnitudes, entries that are smaller will suffer from higher relative error.  In the case of\anonymized{Product A}{EIE}, this can occur because for a given trip, the distance traveled by walking is typically on the order of kilometers, while the distance traveled by flying can be on the order of hundreds or thousands of kilometers.  Additionally, values for different metrics have different units, and can also vary dramatically (e.g., distance is measured in kilometers, while duration is measured in seconds).

\item  \texttt{Budget Split.} Another way to apply the Laplace mechanism is to split the privacy budget among some subset of groups and metrics, and run multiple independent copies of the Laplace mechanism for each ``slice`` of the joint histogram.  This can get around the issue of the \texttt{Joint Clipping} baseline since the clipping norm for each slice of the joint histogram can be calibrated to the magnitude of the entries in that slice, providing more uniform relative errors across the joint histogram.  The drawback of this approach is that it requires splitting the budget across slices of the histogram, and therefore the noise added to each entry is potentially significantly larger.  In the case of\anonymized{Product A}{EIE}, it would be most natural to split the privacy budget $27$ ways among the $3$ metrics and $9$ activities, as magnitudes for these slices can vary dramatically.

\item \texttt{Group Scaling + Joint Clipping.} Our approach overcomes the limitations of the baseline approaches described above.  Our key idea is to use the \texttt{Joint Clipping} approach, but carefully scale the entries of $\mathbf{v}_i$ prior to clipping so that each slice of the histogram has similar magnitude.   After adding Laplace noise in this transformed domain, we invert the scaling to bring the entries of the noisy histogram back to the right magnitude.  The main technical challenge to solve is how to find good scaling parameters.  A natural choice is to simply use the inverse clipping parameters that would have been used by the \texttt{Budget Split} baseline.
\end{enumerate}

After noising, we also recommend discarding partitions whose noised values fall under a threshold, to eliminate partitions with very little data.

\begin{example}[Activity + Metric Scaling]
For\anonymized{Product A}{EIE}, it is natural to instantiate the \texttt{Group Scaling + Joint Clipping} mechanism across the \activity and \metric dimensions as magnitudes of user contributions for these slices can vary dramatically.  In production, we compute the scaling factors for each (\activity, \metric) pair by computing the appropriate quantiles on a server-side proxy dataset that was available during migration to \pap.  If such a proxy dataset is not available, one can use the exponential mechanism to estimate the quantile with DP instead.

This mechanism is particularly effective for the\anonymized{Product A}{EIE} application because the outlier users for each \activity tend to be different.  That is, there may be many users who are extreme cyclists, runners, flyers, drivers, etc., but few users who are extreme in multiple categories.  However, across metrics, this is less true as e.g., duration and distance traveled are highly correlated.  Hence, there is less to be gained by considering the the metrics jointly instead of simply splitting the privacy budget across them.
\end{example}

Our \texttt{Activity+Metric Scaling Mechanism} is summarized in \cref{fig:mechanism} and described more precisely below, as well as in \cref{alg:clientwork,alg:serverwork}.  We will use the notation $\mathbf{v}(a, m, r, d)$ to index into a histogram defined over activities $a$,  metrics $m$, regions $r$, and directions $d$.

\begin{enumerate}
    \item Let $S(a, m)$ be the 95\% quantile of $|| \mathbf{h}_i(a, m, \cdot, \cdot) ||_1$, where $\mathbf{h}_i(a, m, \cdot, \cdot)$ is the subvector of $\mathbf{h}_i$ corresponding to activity $a$ and metric $m$.  We computed these scale values over server-side proxy data, but in principle they could be computed with DP + federation on the on-device data.
    \item Scale each user contribution: Now let $\mathbf{v}_i(a, m, r, d) = \nicefrac{\mathbf{h}_i(a, m, r, d)}{ S(a, m) }$.
    \item Invoke the Laplace mechanism primitive (joint clipping, aggregation, and noise) on $\mathbf{v}_1, \cdots, \mathbf{v}_n$ to get $\tilde{\mathbf{v}}$.
    \item Apply the reverse scaling and return $\tilde{\mathbf{h}}(a, m, r, d) = \tilde{\mathbf{v}}(a, m, r, d) S(a, m)$.
\end{enumerate}

%% file: sections/dp_queries.tex
\begin{figure}[h]
\centering
\begin{minipage}{0.45\textwidth}
\begin{algorithm}[H]
\caption{ClientWork} \label{alg:clientwork}
\begin{algorithmic}
\small
\Require scales $S \in \mathbb{R}^{9 \times 3}$, clip $C \in \mathbb{R}$, user $i \in \mathbb{N}$ 

\State $\mathbf{v} = \mathbf{0} \in \mathbb{R}^{9 \times 3 \times 50K \times 3}$
\For{activity $a$, region $r$, direction $d$, distance, duration \textbf{in} records$_i$}
    \State $\mathbf{v}(a, \text{`num trips'}, r, d) \mathrel{+}= $ $\frac{1}{S(a, \text{`num trips'})}$
    \State $\mathbf{v}(a, \text{`distance'}, r, d) \mathrel{+}= $ $\frac{\text{distance}}{S(a, \text{`distance'})}$
    \State $\mathbf{v}(a, \text{`duration'}, r, d) \mathrel{+}= $ $\frac{\text{duration}}{S(a, \text{`duration'})}$
\EndFor
\State \Return $\mathbf{v} \cdot \min(\frac{C}{|| \mathbf{v} ||_1}, 1)$
\State 
\State
\end{algorithmic}
\end{algorithm}
\end{minipage}
\hspace{2em}
\begin{minipage}{0.45\textwidth}
\begin{algorithm}[H]
\caption{ServerWork} \label{alg:serverwork}
\begin{algorithmic}
\small
\Require scales $S \in \mathbb{R}^{9 \times 3}$, clip $C \in \mathbb{R}$, privacy budget $\epsilon$

\For{User $i$}
    \State  $\mathbf{v}_i =$ ClientWork($S$, $C$, $i$) \Comment \textcolor{orange}{\textbf{Client}}
\EndFor
\State $\mathbf{v} =$ CrossDeviceSum($\mathbf{v}_i$) \Comment{\textcolor{orange}{\textbf{Client $\rightarrow$ Server}}}
\State $\tilde{\mathbf{v}} = \mathbf{v} + \text{Lap}\Big(\frac{C}{\epsilon} \Big)^{9 \times 3 \times 50K \times 3}$ \Comment \textcolor{orange}{\textbf{Server}}

\For{\textbf{each} activity $a$, metric $m$, region $r$, direction $d$}
    \State $\tilde{\mathbf{h}}(a, m, r, d) = \tilde{\mathbf{v}}(a, m, r, d) S(a, m)$  \Comment \textcolor{orange}{\textbf{Server}}
\EndFor
\State \Return $\tilde{\mathbf{h}}$
\end{algorithmic}
\end{algorithm}
\end{minipage}
\end{figure} 

%% file: sections/eval_intro.tex
We evaluate our system on a production use case in terms of (1) device reach, (2) accuracy, and (3) on-device resource consumption. This allows us to study the utility and bias of the results at scale \textemdash how effective it is in establishing sufficient coverage to reach a representative sample of devices \textemdash and quantify resource consumption tradeoffs for participant devices. In deploying our system, we validated that it scales to hundreds of millions of clients and billions of focused updates per day. We also measure the error introduced by the differential privacy mechanisms and compare our DP method to other baseline methods to show how the relative error varies across different privacy budgets. We show that our method allows for reasonable privacy while maintaining high utility for the sustainability application.

{\bf Accuracy Metrics.} We use two different metrics for the FA and DP components of our system in terms of the error they introduce. As described in more detail in \cref{sec:dp_eval}, we judge our DP results using\textit{ weighted relative error}.  However, because this metric heavily weights buckets with many participating devices, we also measure the per-user mean error introduced by the FA infrastructure. This computes the relative error introduced between server-side results, and normalizes by the number of client devices contributing to each partition.

%% file: sections/scale_1.tex

{\bf Measuring device reach.} We assess the systems ability to obtain accurate statistics with minimal bias in terms of device reach. Since the total number of devices in the target population $P$ is unknown\footnote{Some devices may be in use without network connectivity, for example.}, we define device reach $H = \frac{|U|}{|F|}$ as the fraction of the cardinality of the subset $U$ of devices successfully contributing data to a query over the cardinality of the subset $F$ of devices that have the feature activated (we obtain this through conventional telemetry---when the feature becomes active, devices report this to the server).

\begin{figure}[h]
  \includegraphics[width=\columnwidth]{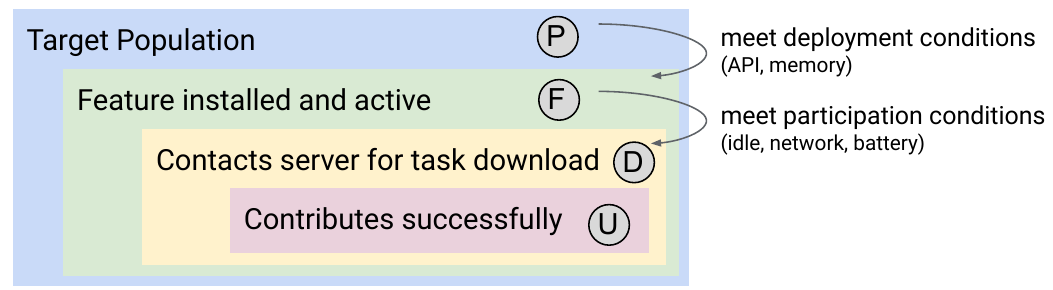}
  \caption{Computation criteria for FA. \label{fig:funnel}}
\end{figure}

 
\cref{fig:funnel} illustrates the device reach funnel. In order to install and activate the feature (subset F), the phone must have at least 1 GB RAM, Android Marshmallow or newer, and occasional network connectivity (this is to eliminate any residual probability of issues on lower-end devices that do not have sufficient resources for this feature). Once active, the feature can either be directly invoked by another module when fresh data becomes available, or it can schedule regular invocations through Android JobScheduler under a set of invocation constraints. 

Query processing can be invoked through some combination of the following conditions: 1) the phone is idle for an extended period; 2) the phone is either connected to an unmetered network (WiFi), or has any connection present; 3) the phone has at least 30\% battery level and is actively charging. When these conditions are met and the feature is invoked, the device attempts to contact the server to signal its availability to execute queries and download available tasks (subset D). It then executes the tasks and, if network connectivity and other constraints permit, reports the results from the on-device queries back to the server (subset U). While other conditions may be flexible, the network connectivity and 30\% battery level requirements remain constant.

{\bf Baseline.} To assess device reach and bias, we use an implementation over the federated learning architecture~\cite{bonawitz2016} as baseline (version 0.1), since this architecture most closely matches the design requirements of supporting flexible on-device computations for early data minimization and immediate aggregation at the server. Note that due to the learning origins of this platform it requires computations to be specified in TensorFlow and the resource needs of the TensorFlow engine warrant a particularly conservative set of invocation constraints (simultaneously WiFi, idle, and charging) to minimize residual risks of phone usage disruptions for users.

{\bf Our Design Iterations.} Version 0.2 of \pap used a built-in SQL-based on-device engine instead of a TensorFlow-based one. Version 0.3 improved the aggregation server by converting the TensorFlow-based (version 0.1) and SQL-based (version 0.2) grouped summation over cross-client contributions to a much more efficient C++ implementation. Given the huge number of devices, we also had to address concerns around task creation and assignment. Specifically, as the data aggregation is synchronous across a cohort of devices, we need to try to accommodate as as many device check-in requests as possible. Otherwise, if many clients attempt to check in to the server to receive an assignment and are turned away, some devices may not able to make a contribution in the future.

We implemented a more efficient task creation and assignment protocol in version 1.0. As part of this protocol, the server can upload multiple tasks to a device during one check-in stage, so the device can compute all of these queries in one wake-up session. If data analysts wish to get different statistics by deploying multiple different queries, they can set all of these queries within one population. The queries will be uploaded to devices simultaneously, addressing the risk of having insufficient eligibility windows for a given device within any time period. Running multiple tasks in a single computation block does run the risk of keeping an Android wakelock ~\cite{wakelock} for too long and draining the battery, but to address this, the client runtime maintains a strict limit of one minute per wake up session. If the limit is exceeded, the runtime terminates all ongoing computations and frees up resources.



{\bf Results.} \cref{tab:meanerr} shows the relative weekly device reach of the proposed system compared to the baseline, as well as ablation results with subsets of the proposed system's features. The weekly device reach is 93\% for the proposed system, compared to 49\% for the baseline. This suggests that results from the proposed system are significantly more representative of the underlying target population. This can also be observed from the per-user mean error shown in the last column. The metric here shows the increase in error relative to the proposed system. For example, with device reach of 84\%, the error increases by 83\% compared to the proposed system with a device reach of 93\%.


\begin{table}
\centering
  \begin{tabular}{p{0.4\linewidth} | P{0.12\linewidth} | p{0.3\linewidth}}
    \textbf{System Version} & \textbf{Relative Device Reach} & \textbf{Per-User Mean Error Ablation} \\
     & (all regions) & (relative to the 1.0 proposed system) \\\hline
     0.1 Baseline & 49\% & * \\
     0.1 Baseline without unmetered network and charging constraints & 76\% & *\\
     0.2 Baseline w/ On-device SQLite engine & 84\% & 83\%\\
     0.3 More efficient aggregation server grouped summation & 89\% & 27.5\%\\
     1.0 Proposed System & 93\% & N/A \\
  \end{tabular}
  \caption{Per-user mean error comparison across the platform versions. (*) No comparable data available: limited experiment due to increased resource consumption. \label{tab:meanerr}}
\end{table}

\cref{tab:reachtable} shows the difference in reach between the baseline invocation constraints (including WiFi and charging for more resource intensive workloads) and our proposed constraint for lighter workloads that only requires idle devices. It further breaks down by different countries, as well as daily and weekly device reach. The comparison shows that our lightweight and latency-tolerant design, which accepts contributions over at least a week, not only improves overall device reach (up to 76.4\% across all countries), but also reduces geographic differences. We attribute the remaining differences to differing network connectivity and usage patterns across countries, as well a different distribution of higher-end versus lower-end devices.


\begin{table}
\vspace{-1em}
\centering
  \begin{tabular}{c|cc}
    \textbf{Constraints} & \multicolumn{2}{c}{\textbf{Device Reach}} \\
     & Weekly & Daily \\\hline
    \textbf{All countries} & & \\
    Idle, WiFi, Charging & 49.4\% & 23.8\% \\
    Idle & 76.4\% & 44.2\% \\\hline
    \textbf{USA} & & \\
    Idle, WiFi, Charging & 48.3\% & 22.7\% \\
    Idle & 74.9\% & 42.0\% \\\hline
    \textbf{India} & & \\
    Idle, WiFi, Charging & 19.8\% & 6.5\% \\
    Idle & 63.9\% & 28.2\% \\\hline
    \textbf{Finland} & & \\
    Idle, WiFi, Charging & 40.6\% & 17.6\% \\
    Idle & 82.1\% & 48.6\% \\
  \end{tabular}
  \caption{Device reach with different conditions with the private analytics platform v.0.1. \label{tab:reachtable}}
  \vspace{-1.5em}
\end{table}



{\bf Device resource efficiency.} For the\anonymized{Product A}{EIE} use case, we measured that for the 95th percentile of all devices, the sum of total download and upload data transferred per query is less than $ 15 kB $, and performing the full computation takes roughly $8.5$ seconds. This includes checking in to the server, waiting for the task to be downloaded, running a query, and uploading results. The query computation itself contributes less than a second to this total time.

Based on these timing and bandwidth measurements and a federated system energy model~\cite{patterson2024}, we estimate that on-device processing and communication of a query consumes less than $3.5J$ when on a cellular connection, and less than $3J$ over WiFi. This represents less than 0.035\% of a typical $3000mAh$ battery. With weekly invocations for the\anonymized{Product A}{EIE} query and the aforementioned system optimizations, this yields an amortized daily battery consumption of less than 0.005\%. Note that this estimate is still likely to be high, since it assumes that the phone initiates the radio transmission solely for this one query. In practice, the connection is often shared with other traffic, thus further amortizing the energy consumption.

%% file: sections/dp_eval.tex
In this section, we evaluate three different mechanisms for privately estimating the\anonymized{Product A}{EIE} histograms: the two baseline approaches described in \cref{sec:mechanism_design}, as well as our novel approach.  These experiments were run on our server-side proxy dataset, which allowed us to compare mechanisms efficiently, and estimate how well they might do with different privacy budgets before deployment.  

\paragraph{\textbf{Accuracy Criteria}}

Given noisy answers to the workload queries defined in \cref{example:workload}, we judge their utility according to a weighted relative error, defined as follows.  For a given metric and each (\region, \direction, \activity) we calculate the relative error between the un-noised aggregate value and noisy estimate.  We then take a weighted average of this quantity across the entire histogram, ignoring entries where fewer than $2000$ client devices contributed data (if the data is too noisy, these entries will go unreleased anyways, but the privacy protections will still apply).  The weight for a given \region $r$, \direction $d$ and \activity $a$ is defined as $n_{r, d, a} / n_r$, where $n$ is the total number of trips for a given setting.  This weights entries according to how important they are within the given region.  

In this work, the end application requires an average relative weighted error of $\approx 3\%$  to be useful in downstream applications.
We vary the privacy budgets, and include a reference line for our target $\approx 3\%$ relative error.

\begin{figure*}
\vspace{-1.5em}
\centering
\hspace{-5em}
\begin{subfigure}{0.45\textwidth}  
    \centering
    \includegraphics[width=\textwidth]{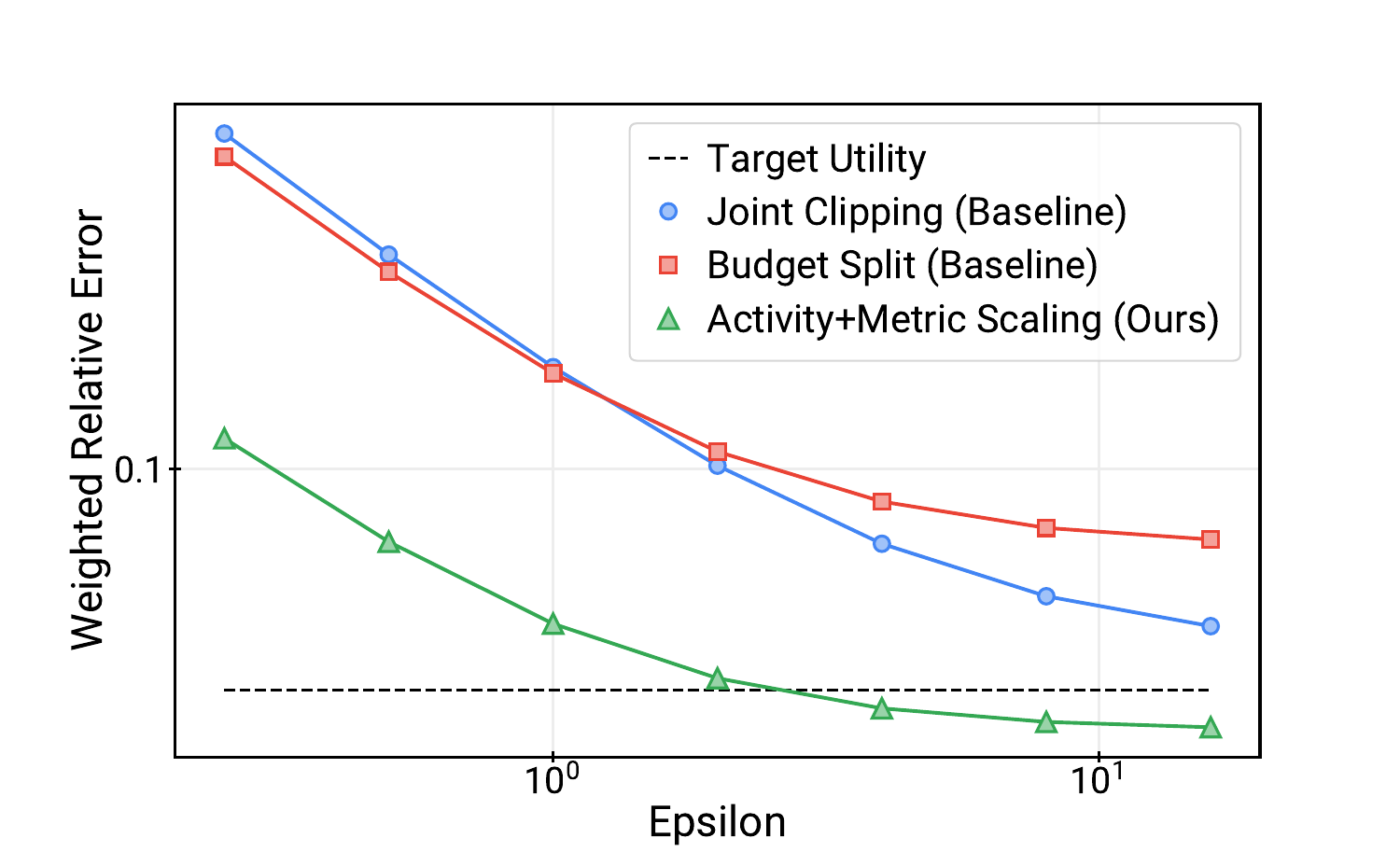}
    \caption{\label{fig:plot}}
\end{subfigure}
\begin{subfigure}{0.45\textwidth}
    \centering
    \vspace{2.5em}
    \begin{tabular}{c|ccc}  
        \textbf{Mechanism} & \multicolumn{3}{c}{\textbf{Weighted Relative Error}} \\
        ($\epsilon = 2$) & Num Trips & Distance & Duration \\\hline
        Joint Clipping & 0.195 & 0.072 & 0.038 \\
        Budget Split & 0.091 & 0.150 & 0.088 \\
        Activity + Metric & 0.028 & 0.040 & 0.028 \\
        Scaling & & & \\
    \end{tabular}
    \vspace{3em}
    \caption{\label{fig:table}}
\end{subfigure}
\caption{(a) overall weighted relative error of two baseline mechanisms and our mechanism for varying privacy budgets.  (b) Error breakdown of each mechanism for each metric at $\epsilon=2$.}
\label{fig:experiments}
\end{figure*}

\paragraph{\textbf{Analysis of Results}}
As shown in \cref{fig:plot}, among the two baselines discussed in \cref{sec:mechanism_design}, the \texttt{Budget Split} baseline was slightly better for smaller privacy budgets, while the \texttt{Joint Clipping} baseline was better for larger privacy budgets.  Neither baseline met the utility goal of $3\%$ weighted relative error even for $\epsilon = 16$.  Our new \texttt{Activity + Metric Scaling} mechanism performed much better across the board, and came close to the utility target for $\epsilon = 2$.  

As shown in \cref{fig:table}, the \texttt{Joint Clipping} baseline enjoyed 5$\times$ lower error on the \duration metric than the \numtrips metric.  This is likely due to the fact that the duration (measured in minutes) of most trips is much greater than $1$, so those histogram entries of the histogram are much larger than those corresponding to \numtrips.  The \texttt{Budget split} baseline has similar error between \duration and \numtrips'' but a 2$\times$ worse error for the \distance metric.  We believe this might be because the distance traveled is a higher variance quantity and we lose significant signal from the outlier users above the 95\% quantile.  Our new mechanism achieves similar error between all three metrics, although the \distance error is still somewhat larger.

%% file: sections/related_work.tex
To give strong privacy guarantees like the ones in Mayfly, prior works have proposed solutions based on local differential privacy (LDP), which involve adding noise at the device level, before aggregating across users. These include deployments by Apple~\cite{apple2017learning}, Microsoft~\cite{ding2017collecting}, Snap~\cite{pihur2022differentially}, and academic works~\cite{acs2011have}. While these systems avoid placing trust in central aggregators, they require adding noise that scales with the number of users, and thus suffer from much worse privacy-utility tradeoffs, especially on the scale of hundreds of millions of users. Previous work from Google entitled RAPPOR~\cite{erlingsson2014rappor, fanti2016building} uses a one-hot encoding scheme to perform its local noising operations, but the resulting operations are intensive in terms of compute, memory, and bandwidth, as the message lengths sent by users are a function of the number of aggregated metrics.  Several solutions have been proposed to reduce the compute/bandwidth requirements of local approaches like these, including sketching~\cite{sun2024private, zhu2020federated, chadha2024differentially, bagdasaryan2022towards} and adaptivity~\cite{mckenna2020workload, acharya2019hadamard}. However, the additional operations in these works make the privacy-utility tradeoffs even less favorable.

To deal with the tradeoffs in the LDP model, shuffling~\cite{erlingsson2019amplification, balle2023amplification, cheu2019distributed} has emerged as a technique to amplify privacy guarantees, with some systems implementing this technique at scale~\cite{bittau2017}. However, achieving close-to-asymptotically optimal accuracy appears to demand a large volume of messages from each user ~\cite{balcer2019separating}. Other works pick different spots on this tradeoff curve (e.g. ~\cite{ghazi2021power, cheu2022differentially}), but all require the presence of a service that applies uniform shuffle; securely implementing that shuffler is a nontrivial task. There are also approaches that involve distributed DP noise generation, although the accuracy of this setting still does not match the central setting.

The works that operate in the same setting as Mayfly are in the central model, where noise is added only after aggregation. There have been previous SQL-based central DP system designs such as PINQ~\cite{mcsherry2009privacy} and Airavat~\cite{roy2010airavat}, but these are server-side only, and don't deal with any privacy guarantees in aggregating client data. To deal with the trust requirement introduced by a central aggregator, there are solutions that rely on a group of servers or elected committee devices to distribute trust, including Honeycrisp~\cite{roth2019honeycrisp} and Unlynx~\cite{froelicher2017unlynx}. PANDA~\cite{wang2022panda} shows how to use TEEs and authenticated encryption to create a lightweight non-interactive private aggregation scheme. This follows other works using cryptographic primitives to implement secure aggregation schemes in a single-server setting. Mansouri et. al~\cite{mansouri2023sok} provide a good overview of these works, but we highlight a few solutions based on homomorphic encryption~\cite{jawurek2012fault}), pairwise masking~\cite{ma2023flamingo}, and secret sharing~\cite{so2022lightsecagg, kadhe2020fastsecagg}. All of these involve additional communication overhead, which makes them difficult to implement across large numbers of devices (many of them potentially with low resources). 

Additional central DP systems that operate on streams in the single-server setting include ~\cite{zhang2024, chen2017pegasus}. These systems are designed for much simpler applications, where we do not have substantially varying signals across users and devices. There is also significant engineering and systems innovation required to transform them into systems that work with large (500M+) numbers of devices. 

Other systems that operate in a similar space but do not natively offer DP guarantees include Prio~\cite{corrigangibbs2017}, which can optionally support DP, but only after aggregation – it offers no support for local contribution bounding to support optimized queries of the kind we propose in Mayfly, and also relies on a group of servers to enforce its trust guarantees.

%% file: sections/discussion.tex
This work reflects experience from deploying privacy-preserving analytics and learning systems over several years. It also points to open questions.

{\bf DP-ephemerality tension.} Based on exploring real-world constraints and privacy risks, we believe that ephemerality offers substantial privacy benefits. Often there are many means (including non-technical ones) for attempting to track someone's future behavior, but comparatively few means for obtaining information about someone's past behavior. An ephemeral system design that does not retain a history of individual user data, therefore significantly reduces the information value that can be obtained from the system and virtually eliminates a large slice of insider risks. However, there exists a tension between achieving a fully ephemeral system design and using conventional DP mechanisms. This is because these mechanisms require bounded sensitivity of user contributions, and offer the best utility-privacy tradeoff when applied over large temporal privacy units (long windows). To effectively bound user contributions over longer windows, some per-user information has to be maintained, at least for the duration of the window. We have addressed this in our design by retaining such information only on-device and only for short windows---a more fundamental solution remains an open problem for future work.

{\bf Device resource and bias tradeoffs.} When designing for worldwide deployment, device resource budgets become especially tight. There is a long tail of low-cost smartphone devices, and a design that aims for unbiased results needs to target a lower percentile in this device distribution, since the usage of more capable devices tends to correlate with higher socioeconomic status. For these target devices, we aim for the total resource usage to remain low, since participating in our system is just one out of many features that might use device resources. We have addressed this through system optimizations and a latency-tolerant design that allows devices to contribute data over WiFi whenever possible, but open challenges remain for more latency-sensitive applications.

%% file: sections/conclusions.tex
The Mayfly platform successfully powers a sustainability query on a scale of hundreds of millions of unique contributing devices, allowing for a $ \epsilon = 2 $ (device,week)-level-DP guarantee while meeting strict utility requirements.
The platform infrastructure introduces a federated architecture, designed not for machine learning but specifically for analytics, supporting synchronous ephemeral aggregations in a streaming model. Limitations in the threat model (the system does not provide rigorous protection against logging or inspecting un-aggregated messages) suggest an opportunity for future work with more enhanced confidentiality. Follow-up work could adopt confidential federated analytics~\cite{eichner2024, vanoverveldt2025discovering}, a novel system architecture based on trusted execution environments. This architecture supports asynchronous uploads and an option to reuse the same uploaded data for multiple workloads, while ensuring confidentiality of server-side computations and providing externally verifiable privacy properties.


%% file: sections/ack.tex
Like most large scale system efforts, there are many more contributors than the authors of this paper. The following people have directly contributed to coordination, design, and implementation: Brett McLarnon, Chloé Kiddon, Chunxiang (Jake) Zheng, Emily Glanz, Katharine Daly, Hubert Eichner, Mira Holford, Nova Fallen, Rakshita Tandon, Stefan Dierauf, Scott Wegner, Zachary Garrett, Ulyana Kurylo, Octavian Suciu, Steve He, Prem Eruvbetine, Emma Barron, Cris Castello, Andres Soto. We would also like to thank Brendan McMahan and Daniel Ramage for helpful comments, discussion, and leadership.

%% file: sections/dp_spec.tex
We give a self-contained description of the DP guarantees of \pap, following the recommended guidelines from \cite{ponomareva2023dp}.

\begin{enumerate}
    \item \textbf{DP setting.} This a central DP guarantee where the service provider is trusted
to correctly implement the mechanism.
    \item \textbf{Instantiating the DP Definition.}
    \begin{enumerate}
        \item \textit{Data accesses covered:} This DP guarantee applies to all participating devices who opt-in to participate in a round of aggregation and upload their data for a given time window.
        \item \textit{Final mechanism output:} For each time window covering a privacy unit, a DP histogram is released to internal data analysts. Some subset of this histogram will also be released to external parties.
        \item \textit{Unit of Privacy:}  We guarantee (device, window)-level privacy. For our EIE case study, we use a window of one calendar week.
        \item \textit{Adjacency definition for ``neighboring'' datasets:} We use the add-remove definition of neighboring, where neighboring data sets differ by addition/removal of a single device for any time window.
    \end{enumerate}
    \item \textbf{Privacy accounting details}
    \begin{enumerate}
        \item \textit{Type of accounting used:} No basic or advanced composition is needed, since we have a single application of the Laplace mechanism.
        \item \textit{Accounting assumptions:} NA
        \item \textit{The formal DP statement:} For our EIE case study, we guarantee DP of $\epsilon=2$ per week.
    \end{enumerate}
    
    \item \textbf{Transparency and verifiability:} Our current implementation is not open-source, but we plan to open-source DP mechanisms in the future.

\end{enumerate}